\documentclass[preprint,showpacs,preprintnumbers,amsmath,amssymb]{revtex4}

\usepackage{graphicx}
\usepackage{dcolumn}
\usepackage{bm}

\begin{document}

\preprint{HE - 02}

\title{Magnetic-field-induced switching between ferroelectric phases in orthorhombic-distortion-controlled $R$MnO$_{3}$}

\author{K. Noda, M. Akaki, T. Kikuchi, D. Akahoshi,}
\author{H. Kuwahara}%
\affiliation{%
Department of Physics, Sophia University\\
Chiyoda-ku, Tokyo 102-8554, Japan}

\date{\today}

\begin{abstract}
We have investigated the dielectric and magnetic properties of Eu$_{0.595}$Y$_{0.405}$MnO$_{3}$ $without$ the presence of the 4$f$ magnetic moments of the rare earth ions, and have found two ferroelectric phases with polarization along the $a$ and $c$ axes in a zero magnetic field. 
A magnetic field induced switching from one to the other ferroelectric phase took plase in which the direction of ferroelectric polarization changed from the $a$ axis to the $c$ axis by the application of magnetic fields parallel to the $a$ axis. 
In contrast to the case of TbMnO$_{3}$, in which the 4$f$ moments of Tb$^{3+}$ ions play an important role in such a ferroelectric phase switching, the magnetic-field-induced switching between ferroelectric phases in Eu$_{0.595}$Y$_{0.405}$MnO$_{3}$ does not originate from the magnetic transition of the rare-earth 4$f$ moments, but from that of the Mn 3$d$ spins. 
\end{abstract}

\maketitle

\newpage 
\newpage


The first observation of the magnetoelectric effect was carried out for Cr$_{2}$O$_{3}$ in 1960.\cite{astrov}
Through the 1960s and 1970s, many magnetoelectric materials were discovered, such as Ti$_{2}$O$_{3}$, $R$AlO$_{3}$, $R$PO$_{4}$ ($R$ is a heavy rare earth ion), and so on.\cite{alshin, holmes, honreich, rado} 
Unfortunately, the magnetoelectric effects in such materials are too small for use in electronic devices. 
Since the recent discovery of the ``\underline{m}agnetic-field-induced \underline{e}lectric \underline{p}olarization \underline{f}lop" (MEPF) in TbMnO$_{3}$,\cite{kimura} the magnetoelectric effect has attracted revived interest. 
Even among the manganese based oxides, several magnetoelectric materials have been found and studied extensively, including GdMnO$_{3}$, Gd$_{1-x}$Tb$_{x}$MnO$_{3}$ (0$<$$x$$<$1), DyMnO$_{3}$, TbMn$_{2}$O$_{5}$, and DyMn$_{2}$O$_{5}$.\cite {kuwahara, noda, goto, hur, higashiyama} 
The relatively large magnetoelectric effect demonstrated in TbMnO$_{3}$ is believed to originate from the strong $fd$ exchange interaction between the Mn 3$d$ spins and the Tb 4$f$ moments and from the magnetic frustration due to the large orthorhombic distortion. 
It is widely accepted also that, in other orthorhombic $R$MnO$_{3}$ compounds, the magnetic transition of the rare-earth 4$f$ moments plays an important role in the magnetoelectric effect. 
However, the origin of the magnetoelectric effect in manganites has not yet been completely accounted for by the above simple scenario. 
Therefore, to ascertain whether the rare-earth 4$f$ moments are essential to produce such a peculiar magnetoelectric effect as MEPF, we have investigated this effect in orthorhombic $R$MnO$_{3}$ $without$ the presence of the 4$f$ magnetic moments of rare earth ions. 
For the $R$-site ion, we chose an admixture of Eu$^{3+}$ and Y$^{3+}$ ions to remove or minimize the influence of the 4$f$ moments; in addition, we controlled the average ionic radius of the $R$ site so that it was the same as that of TbMnO$_{3}$. 
The Eu$^{3+}$ ion is thought to have no 4$f$ moments because of the fact that the total angular momentum $J$=0, while the Y$^{3+}$ ion does not have 4$f$ electrons. 
The subject compound, Eu$_{0.595}$Y$_{0.405}$MnO$_{3}$ ((Eu,Y)MnO$_{3}$), was, therefore, expected to be free from 4$f$ moments.  
Our results revealed a magnetic-field-induced switching between ferroelectric phases in (Eu,Y)MnO$_{3}$. 
In this paper, we discuss the role of Mn 3$d$ spins in such a phase switching in this compound.


We prepared Eu$_{0.595}$Y$_{0.405}$MnO$_{3}$, whose average ionic radius of the $R$ site is the same as that of TbMnO$_{3}$. 
The compositional ratio of the Eu$^{3+}$ to Y$^{3+}$ ions was obtained by a calculation based on the Shannon's ionic radius table.\cite{shanon} 
The single crystal sample was grown by the floating zone method. 
We performed x-ray-diffraction and rocking curve measurements on the resulting crystal at room temperature, and confirmed that it showed the single phase with the orthorhombic $Pbnm$ structure without any impurity phases such as hexagonal one or any phase segregation. 
All specimens used in this study were cut along the crystallographic principal axes into a rectangular shape by means of an x-ray back-reflection Laue technique. 
The measurements of the temperature dependence of the dielectric constant and the spontaneous ferroelectric polarization in magnetic fields were carried out in a temperature-controllable cryostat equipped with a superconducting magnet that provided a field up to 8T\@.
The dielectric constant was determined with an LCR meter (Agilent, 4284A). 
After the sample had been cooled under a poling electric field of 300$\sim$500 kV/m, the spontaneous electric polarization was obtained by the accumulation of a pyroelectric current while it was heated at a rate of 4K/min. 
The magnetization and specific heat were measured using a commercial apparatus (Quantum Design, PPMS).


We show in Fig.\ \ref{fig1} the temperature dependence of the dielectric constant along the $a$ ($\varepsilon_{a}$) and $c$ ($\varepsilon_{c}$) axes [(a),(d)], and the spontaneous ferroelectric polarization along the $a$ ($P_{a}$) and $c$ ($P_{c}$) axes [(b),(e)]. 
In addition, we include the magnetization (c) in magnetic fields parallel to the $a$ axis ($H_{a}$), and the temperature dependence of the magnetization in the field parallel to each axis (f). 
First, we will discuss the overall magnetic behavior in a magnetic field of 0.5T. 
As clearly seen in Fig.\ \ref{fig1} (c), the normalized magnetizations in 0.5T and 1T are almost coincident, a result which indicates that, below 1T\@ , the magnetization changes linearly with the magnetic field.
Therefore, the temperature dependence of the magnetization shown in Fig.\ \ref{fig1} (f) follows almost the same trend as that in a zero magnetic field. 

Other $R$MnO$_{3}$ crystals related to this study are known to have an anisotropic magnetic structure: either a layered ($A$-type) or a commensurate/incommensurate antiferromagnetic structure.\cite{kimura3, kajimoto, kimura4, kuwahara, goto} 
The present compound (Eu,Y)MnO$_{3}$ also shows a remarkable anisotropic magnetization below 47K\@, above which temperature a common paramagnetic behavior is observed along all axes. 
Below 47K\@, the magnetization parallel to the $b$ axis steeply decreases, while in the temperature range from 47K to 25K, the magnetization along the $a$ ($M_{a}$) and $c$ ($M_{c}$) axes remains almost constant. 
This result implies that the antiferromagnetic order in which the $b$ axis is a magnetic easy axis appears below 47K\@. 
At 25K\@, a very subtle decrease in $M_{a}$ and $M_{c}$ is detected. 
With a further decrease in temperature, $M_{a}$ and $M_{c}$ shows the opposite behavior below 23K\@: the former fell, and the latter rose again. 

These results suggest that (Eu,Y)MnO$_{3}$ has three magnetic transitions: (1) a paramagnetic to antiferromagnetic (perhaps something like a layered antiferromagnetic) transition at 47K\@; (2) an unknown magnetic transition at 25K, corresponding to the peak in $\varepsilon_{c}$, discussed later; and (3) a canted antiferromagnetic transition with weak ferromagnetism along the $c$ axis at 23K\@, where the Mn 3$d$ spins along the $b$ axis tilt toward the $c$ axis. 
Furthermore, as expected, the 4$f$ moments of Eu$^{3+}$ seem to be quenched at low temperatures in this compound, because the paramagnetic rise and/or anomaly in magnetization, due to 4$f$ moments observed in other $R$MnO$_{3}$ compounds, has not been discerned here. 
The results of specific-heat measurements also support this conclusion: the Schottky anomaly arising from the split levels of the 4$f$ multiplet, also is not detected at low temperatures. 

Now, we turn to a discussion of the temperature dependence of $\varepsilon_{a}$ and $\varepsilon_{c}$ [Figs.\ \ref{fig1} (a) and (d)]. 
Below 2T, $\varepsilon_{a}$ shows a large peak around 23K in the $H_{a}$. 
In the case of $H_{a}$$=$3T, the peak is lowered and broadened, and is shifted to a lower temperature of 19K\@; in addition, and a new anomaly appears around 25K\@. 
Furthermore, in $H_{a}$$\geq$6T\@, the peak disappears and the anomaly observed in 3T remains as a small inflection around 25K\@. 
On the other hand, in the case of $\varepsilon_{c}$, two peaks are observed in a zero magnetic field: in $H_{a}$, the small peak at 23K in 0T shifts toward lower temperatures, and disappears, in a manner similar to the case of the $\varepsilon_{a}$ peak, while the sharp peak at 25K in 0T shifts toward a little higher temperatures. 
The behavior of these two $\varepsilon_{c}$ peaks is in good agreement with the peak and anomaly (inflection) of $\varepsilon_{a}$. 
The temperatures where the two $\varepsilon_{c}$ peaks are observed in a zero magnetic field correspond to the magnetic transition temperatures; the peak at the higher temperature of 25K corresponds to (2), and the other, at the lower temperature of 23K, to (3). 

Next, we focus on the results pertaining to the spontaneous ferroelectric polarization in $H_{a}$. 
Figures \ref{fig1} (b) and (e) show the temperature dependence of $P_{a}$ and $P_{c}$, respectively. 
In a zero magnetic field, $P_{a}$ emerges at 23K, and $P_{a}$ (and also $P_{c}$) can be reversed by a DC electric field, results which provide clear evidence for a ferroelectric phase transition. 
By application of $H_{a}$, the ferroelectric phase with $P_{a}$ (FE$_{a}$) is suppressed toward lower temperatures and finally disappears. 
The transition temperature of $P_{a}$ corresponds well with the peak position of $\varepsilon_{a}$ (or with the small broad peak of $\varepsilon_{c}$). 
On the other hand, in the case of $P_{c}$, another ferroelectric phase with $P_{c}$ (FE$_{c}$) exists only between 23K and 25K in a zero magnetic field. 
With increasing $H_{a}$, FE$_{c}$ is expanding toward lower temperatures and finally FE$_{c}$ becomes dominant at the lowest temperature, concomitantly with the suppression of FE$_{a}$ ($H$$\geq$4T); these findings indicate that a magnetic-field-induced switching from FE$_{a}$ to FE$_{c}$ occurred. 
The FE$_{c}$ also extends toward a little higher temperatures, and the transition temperatures for the appearance and disappearance of $P_{c}$ coincide, respectively, with the sharp and broad peak positions of $\varepsilon_{c}$ (or with the small inflection and broad peak positions of $\varepsilon_{a}$). 
The results obtained mean that first FE$_{c}$ appears at 25K, and then, below 23K, FE$_{c}$ is replaced by FE$_{a}$,  in a zero magnetic field (See also Fig.\ \ref{fig3}). 
The ferroelectric transition temperatures for FE$_{c}$ and FE$_{a}$ are in accord with the magnetic transition temperatures of (2) and (3), respectively. 
In the case of $H_{a}$$\neq$0, FE$_{c}$ develops toward lower temperatures, while FE$_{a}$ is suppressed with an increasing $H_{a}$. 
The transition temperature of the phase switching between FE$_{a}$ and FE$_{c}$ in $H_{a}$ also agrees well with the magnetic transition temperature, as clearly demonstrated in Fig.\ \ref{fig1} (c). 
We have found that the succession of ferroelectric transitions in a zero magnetic field with decreasing temperature is as follows: the paraelectric phase$\rightarrow$FE$_{c}$$\rightarrow$FE$_{a}$. 
Moreover, the FE$_{a}$ and FE$_{c}$ can be controlled by the application of $H_{a}$. 

In order to further investigate the phase switching between FE$_{a}$ and FE$_{c}$ caused by the application of $H_{a}$, we performed isothermal measurements of $P_{a}$, $P_{c}$, and $M_{a}$ as a function of the magnetic field strength (Fig.\ \ref{fig2}). 
The crossover from $P_{a}$ to $P_{c}$ in $H_{a}$ is clearly demonstrated in Figs.\ \ref{fig2} (a) and (b): for example, at 5K, $P_{a}$ abruptly drops around 4T where $P_{c}$ rapidly rises up. 
The magnetic-field-induced switching between FE$_{a}$ and FE$_{c}$ has a certain finite phase-boundary at about 0.5T, in which $P_{a}$ and $P_{c}$ coexist (See also Fig.\ \ref{fig3}). 
As shown in Fig.\ \ref{fig2} (c), $\Delta$$M_{a}$/$\Delta$$\mu$$_{0}$$H_{a}$ also displays a striking peak around the switching field corresponding to the phase boundary. 
The value of $\Delta$$M_{a}$/$\Delta$$\mu$$_{0}$$H_{a}$ shows a step-like increase as $H_{a}$ rose below and above the switching field, which indicates that the direction of the weak ferromagnetic moments (the field-induced canting component) changed from the $c$ axis to the $a$ axis in $H_{a}$. 
These results suggest that the increase of the magnetic moment parallel to the $a$ axis is closely related to the expansion of FE$_{c}$ in higher magnetic fields. 

We present in Fig.\ \ref{fig3} the magnetoelectric phase diagram obtained for (Eu,Y)MnO$_{3}$ in the temperature and magnetic field plane. 
The switching demonstrated between the ferroelectric phases in (Eu,Y)MnO$_{3}$ originates from the magnetic transition of Mn 3$d$ spins alone in sharp contrast to the magnetoelectric effect found in other ferroelectric manganites such as TbMnO$_{3}$ in which the 4$f$ moments are essential. 
Moreover, the MEPF in TbMnO$_{3}$ emerges at the magnetic transition temperature of the Tb 4$f$ moments and shifts toward higher temperatures with increasing magnetic field. However, the observed phase switching in (Eu,Y)MnO$_{3}$ occurs at the canted-antiferromagnetic-transition temperature of the Mn 3$d$ spins and shifts toward lower temperatures. 
This fact also reflects the difference between the magnetoelectric effects $with$ and $without$ the 4$f$ moments.

In summary, we have studied the dielectric and magnetic properties of a Eu$_{0.595}$Y$_{0.405}$MnO$_{3}$ single crystal. 
We have found that this compound has two distinct ferroelectric phases that have $P_{a}$ ($T$$\leq$23K) and $P_{c}$ (23K$\leq$$T$$\leq$25K) in a zero magnetic field. 
In addition, we have demonstrated a magnetic-field-induced switching between these ferroelectric phases. 
In the present case, the direction of ferroelectric polarization is changed from the $a$ axis to the $c$ axis by application of $H_{a}$, a switch which is in the quite $opposite$ direction in the case of TbMnO$_{3}$. 
In contrast to TbMnO$_{3}$ or other multiferroic manganites, (Eu,Y)MnO$_{3}$ does not have 4$f$ moments; therefore, the observed magnetic-field-induced switching between ferroelectric phases is attributed to the magnetic transition of the Mn 3$d$ spins alone. 
The microscopic origin of this ferroelectric phase transition has not yet been clarified. 
However, the observed successive ferroelectric phase transition in (Eu,Y)MnO$_{3}$ $without$ 4$f$ magnetic moments should provide an improved understanding of the mechanism of the magnetoelectric effect not only in manganites but also in other multiferroic materials.


\newpage

\newcounter{volume}
\setcounter{volume}{99}

\newpage

\begin{figure}
\vspace{-3mm}
\caption{Temperature dependence of dielectric constant along the $a$ (a) and $c$ (d) axes and ferroelectric polarization along the $a$ (b) and $c$ (e) axes in magnetic fields parallel to the $a$ axis. Part (c) shows the temperature dependence of the magnetization parallel to the $a$ axis normalized by the magnetic field intensity. The temperature dependence of the magnetization parallel to each axis in a field of 0.5T is shown in (f). For clarity, only the data in a warming scan are plotted.}
\label{fig1}
\end{figure}

\begin{figure}
\vspace{-3mm}
\caption{Magnetic field dependence of ferroelectric polarization along $a$ (top panel) and $c$ (middle) axes at several fixed temperatures. The bottom panel shows the gradient of the $M$-$H$ curve. Magnetic fields are applied parallel to the $a$ axis. For clarity, only the data in the field-decreasing scan are plotted. }
\label{fig2}
\end{figure}

\begin{figure}
\vspace{-3mm}
\caption{The magnetoelectric phase diagram for Eu$_{0.595}$Y$_{0.405}$MnO$_{3}$ in magnetic fields parallel to the $a$ axis.}
\label{fig3}
\end{figure}


\begin{references}

\bibitem{astrov} D. N. Astrov $et$ $al$., Sov. Phys. --JETP {\bf 11}, 708 (1960).

\bibitem{alshin} B. I. Alshin $et$ $al$., Sov. Phys. --JETP {\bf 17}, 809 (1963).

\bibitem{holmes} L. M. Holmes $et$ $al$., Phys. Rev. {\bf B 5}, 147 (1972).

\bibitem{honreich} R. M. Hornreich $et$ $al$., Phys. Rev. {\bf B 16}, 1112 (1977).

\bibitem{rado} G. T. Rado $et$ $al$., Phys. Rev. Lett. {\bf 23}, 644 (1969).

\bibitem{kimura} T. Kimura $et$ $al$., Nature {\bf 426}, 55 (2003).

\bibitem{kuwahara} H. Kuwahara $et$ $al$., Physica {\bf B 359-361}, 1279 (2005).

\bibitem{noda} K. Noda $et$ $al$., IEEE Trans. Magn. (in press; cond-mat/0412148).

\bibitem{goto} T. Goto $et$ $al$., Phys. Rev. Lett. {\bf 92}, 257201 (2004).

\bibitem{hur} N. Hur $et$ $al$., Nature {\bf 429}, 392 (2004).

\bibitem{higashiyama} D. Higashiyama $et$ $al$., Phys. Rev. {\bf B 70}, 174405 (2004).

\bibitem{shanon} R. D. Shannon, Acta Crystallogr. Sect. {\bf A 32}, 751 (1976).

\bibitem{kimura3} T. Kimura $et$ $al$., Phys. Rev. {\bf B 67}, 180401 (2003).

\bibitem{kajimoto} R. Kajimoto $et$ $al$., Phys. Rev. {\bf B 70}, 012401 (2004).

\bibitem{kimura4} T. Kimura $et$ $al$., Phys. Rev. {\bf B 71}, 224425 (2005).

\end{references}
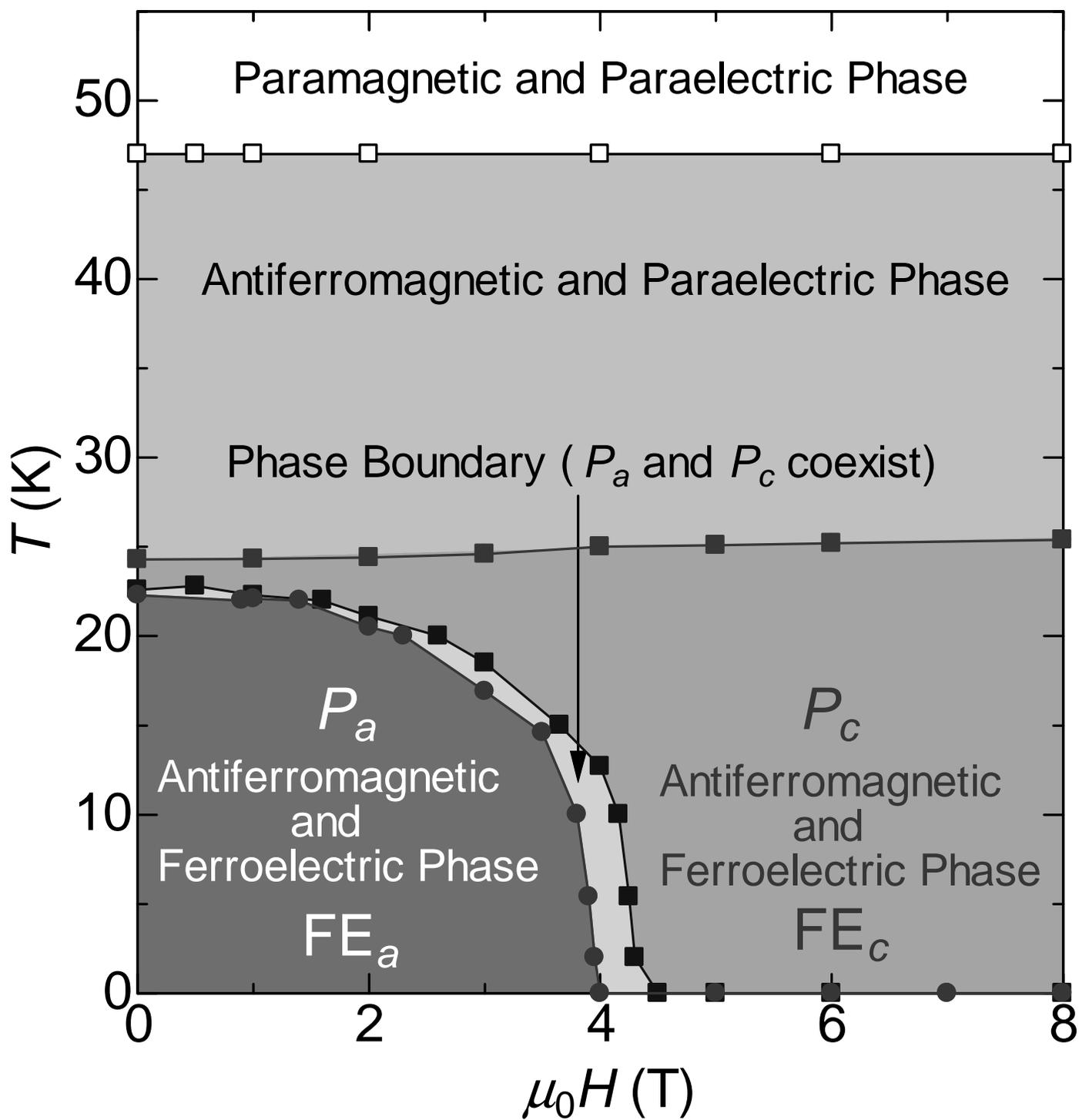
\end{document}